\documentclass{elsart}

\usepackage{epsfig}

\usepackage{amssymb}

\begin{document}

\begin{frontmatter}



\title{Production of $\omega$ mesons in proton-proton collisions\thanksref{BJ}}
\thanks[BJ]{This work was sponsored in part by the BMBF and the Forschungszentrum J\"ulich.}
\center{\large The COSY-TOF collaboration}

\author[5] {S.~Abd El-Samad},
\author[5] {M.~Abdel-Bary},
\author[3] {K.-Th.~Brinkmann \corauthref{c1}},
\corauth[c1]{Technische Universit\"at Dresden, 
Institut f\"ur Kern- und Teilchenphysik, D-01062 Dresden, Germany} 
\ead{KT.Brinkmann@physik.tu-dresden.de}
\author[8] {H.~Clement},
\author[6] {S.~Dshemuchadse}, 
\author[2] {H.~Dutz}, 
\author[4] {W.~Eyrich}, 
\author[8] {A.~Erhardt}, 
\author[5] {D.~Filges}, 
\author[7] {A.~Filippi},
\author[3] {H.~Freiesleben}, 
\author[4] {M.~Fritsch}, 
\author[4] {J.~Georgi},
\author[5] {A.~Gillitzer},
\author[5] {D.~Hesselbarth}, 
\author[3] {B.~Jakob},
\author[3] {L.~Karsch}, 
\author[5] {K.~Kilian}, 
\author[1] {H.~Koch}, 
\author[8] {J.~Kre\ss}, 
\author[3] {E.~Kuhlmann},
\author[7] {S.~Marcello}, 
\author[5] {S.~Marwinski},
\author[1] {S.~Mauro},
\author[1] {W.~Meyer}, 
\author[6] {P.~Michel}, 
\author[6] {K.~M\"oller},
\author[4] {H.~M\"ortel},
\author[5] {H.P.~Morsch}, 
\author[6] {L.~Naumann},
\author[3] {Ch.~Plettner},
\author[3] {M.~Richter},
\author[5] {E.~Roderburg}, 
\author[6] {A.~Schamlott}, 
\author[3] {P.~Sch\"onmeier},
\author[3] {M.~Schulte-Wissermann},
\author[4] {W.~Schroeder},
\author[4] {F.~Stinzing}, 
\author[1] {M.~Steinke},
\author[3] {G.Y.~Sun},
\author[8] {G.J.~Wagner}, 
\author[4] {M.~Wagner},
\author[1] {A.~Wilms},
\author[4] {S.~Wirth}

\address[3]{Institut f\"ur Kern- und Teilchenphysik, Technische~Universit\"at~Dresden, D-01062 Dresden, Germany}
\address[1]{Institut f\"ur Experimentalphysik, Ruhr-Universit\"at~Bochum, D-44780 Bochum, Germany}
\address[2]{Physikalisches Institut, Universit\"at~Bonn, D-53115 Bonn, Germany}
\address[4]{Physikalisches Institut, Universit\"at~Erlangen-N\"urnberg, D-91058 Erlangen, Germany}
\address[5]{Institut f\"ur Kernphysik, Forschungszentrum J\"ulich, D-52425 J\"ulich, Germany}
\address[6]{Institut f\"ur Kern- und Hadronenphysik, Forschungszentrum Rossendorf, D-01314 Dresden, Germany}
\address[7]{INFN Torino, 10125 Torino, Italy}
\address[8]{Physikalisches Institut, Universit\"at~T\"ubingen, D-72076 T\"ubingen, Germany}

\begin{abstract}
The cross section for the production of $\omega$ mesons in proton-proton
collisions has been measured in a previously unexplored region of incident
energies. Cross sections of $\sigma$~=~(7.5\,$\pm$\,1.9)\,$\mu$b and 
$\sigma$~=~(30.8\,$\pm$\,3.4)\,$\mu$b (with 20\% systematic uncertainties)
were extracted at $\epsilon$~=~92~MeV and 173~MeV excess energy 
above the $\omega$ threshold, respectively. The angular 
distribution of the $\omega$ at $\epsilon$~=~173~MeV is strongly anisotropic, 
demonstrating the importance of partial waves beyond pure s-wave production 
at this energy.
\end{abstract}

\end{frontmatter}

The production of mesons in nucleon-nucleon reactions is 
currently being investigated both experimentally and theoretically
because of its implications for the understanding of the nuclear force. In the
early nineties, the comparison of theoretical 
predictions to newly emerging experimental results on near-threshold 
meson production, in particular on pp$\pi$$^0$ \cite{Mey} and, to a lesser 
extent, also other channels, revealed the rather poor understanding of the 
reaction mechanism of inelastic collisions between nucleons. 
Since then, a wealth of
data has been accumulated for a number of mesons in the pseudoscalar sector, 
e.\,g. $\pi$, $\eta$, $\eta$', and K. Recently, investigations in the vector meson
sector ($\rho$, $\omega$, $\phi$) have come into focus. The exchange of these
mesons dominates the nucleon-nucleon interaction at short distances.\\
The comparison of production cross sections of $\omega$ to that of $\phi$ mesons 
may shed light on a possible contribution of strange quark-antiquark pairs in
the nucleon. 
While the $\phi$ contains almost only $s\bar{s}$, the $\omega$ wave function 
is practically decoupled from the strange sector, consisting of  
$u\bar{u}$\,$\oplus$\,$d\bar{d}$ only. The $\phi$ 
production should therefore be strongly suppressed, as predicted 
by the OZI rule \cite{OZI}. A small
contribution of 4$\cdot$10$^{-3}$ is expected because the mixing angle, 
$\Theta$$_V$~$\approx$~39$^{\circ}$ \cite{PDG},
deviates by 3.7$^{\circ}$ from that of a perfectly decoupled system in SU(3), 
$\Theta$$_V$~=~35.3$^{\circ}$.\\
Recently, the DISTO collaboration found that the $\phi$/$\omega$ production
exceeds the OZI prediction by about one order of magnitude \cite{DIS98}. 
Their published data, however, were measured 
at one beam momentum setting of Pc~=~3.67~GeV. Thus, the available 
kinetic energy in the overall CoM system (often referred to as excess energy, 
$\epsilon$) is $\epsilon$~=~85~MeV above the $\phi$ threshold but 
320~MeV for the much lighter pp$\omega$ system. 
The extrapolation down to 85~MeV which 
covers more than one order of magnitude in cross section is far from 
straightforward. It is guided only by reference data measured below $\epsilon$~=~30~MeV
\cite{Hib}. Measurements of $\omega$ production between $\epsilon$~=~30~MeV and
300~MeV are crucial for a more detailed investigation of the excitation
function and, thus, OZI violation in pp collisions.\\
While the $\phi$ angular distribution measured by the DISTO collaboration
\cite{DIS98} is isotropic and therefore compatible with mere s-wave production,
the $\omega$ distribution shows strong contributions from higher partial waves.
If these persist at lower energies, as is the case in one of the possible
scenarios in a recent theoretical calculation of Nakayama and coworkers
\cite{Nak98}, a quantitative comparison of $\phi$ and $\omega$ production may not be
possible without a more detailed understanding of the underlying
(and perhaps different) reaction mechanisms. Hence, experiments should aim at
exclusive differential observables at lower excess energies.\\
In a first attempt to contribute to the issue of $\omega$ production in
proton-proton collisions, data from
an experiment at the Time-of-Flight spectrometer TOF located at an external beam
line of the COoler SYnchrotron COSY at the Forschungszentrum J\"{u}lich were
analyzed for a signal from $\omega$ production. The experimental setup was
described in detail elsewhere \cite{KTB}. The proton beam extracted from the
synchrotron impinges on a thin liquid hydrogen target \cite{Has}. In the present
case, beam momenta of Pc~=~2.95~GeV and 3.2~GeV are used. The reaction products
leaving the target into the forward hemisphere generate, in a thin scintillator, 
a start signal for the successive time-of-flight measurement, then penetrate a
silicon detector and two scintillating-fiber hodoscopes which permit an accurate
track reconstruction needed for the detection of sequential hyperon decay
\cite{pKL}. After flight paths of more than 3~m in vacuum the particles are 
detected in a scintillator assembly with high angular coverage and good spatial 
and time resolution which consists of a barrel section \cite{BAR} and a forward 
structure closing the cylindrical detector at small angles \cite{QUI}. The
detector covers polar angles from about 1$^{\circ}$ up to 70$^{\circ}$.\\
The velocity vectors of charged particles thus measured were used in the 
\protect{analysis} presented here. Events with four hits were selected for further 
inspection. A geometrical separation was applied in order to enrich events of 
the type pp$\pi$$^+$$\pi$$^-$, where the pions may originate from the decay of 
an intermediate resonance, e.g. $\eta$\,$\rightarrow$\,$\pi$$^+$$\pi$$^-$$\pi$$^0$ 
(BR 23\%), $\rho$$^0$\,$\rightarrow$\,$\pi$$^+$$\pi$$^-$ (BR 100\%) or 
$\omega$\,$\rightarrow$\,$\pi$$^+$$\pi$$^-$$\pi$$^0$ (BR 89\%), requiring two 
tracks with angles $\theta$~$\leq$~25$^{\circ}$ (assumed to be protons) and two 
more hits at larger angles (corresponding to the charged pions). This particle 
assignment is justified by kinematics which confines protons 
from pp$\omega$ reactions to the forward cone, while the much lighter pions 
can reach larger angles. The proton identification was further improved by a 
condition on velocities which rejects particles with velocities in excess of that 
of the beam. The proton four-momenta were used to calculate the 
missing mass of a hypothetical third particle, X. If X was a $\rho$$^0$, which 
exclusively decays into two charged pions, both pion directions can be combined 
to give the four-momentum of X. This also holds in case of non-resonant two-pion 
production where the four-momentum of X reconstructed from the missing-mass 
analysis of ppX and that calculated with the X$\rightarrow$$\pi\pi$ hypothesis 
are required to agree within the detector resolution. In case of $\omega$ decay 
into three pions, a $\pi$$^0$ will be missing so that the detected pion pair will 
in most cases
not be consistent with a two-body decay of X. 
This is demonstrated in figure 1, where spectra of the angle of acoplanarity 
between the pion directions with respect to the decaying meson are shown for the 
data (large frame) and for simulated $\rho$$^0$\,$\rightarrow$\,$\pi$$^+$$\pi$$^-$, 
$\eta$\,$\rightarrow$\,$\pi$$^+$$\pi$$^-$$\pi$$^0$ and
$\omega$\,$\rightarrow$\,$\pi$$^+$$\pi$$^-$$\pi$$^0$ events. The acoplanarity was 
defined as the deviation of the angle between the momenta of the pions taken in the
rest frame of the decaying meson from 180$^{\circ}$ (i.e. back-to-back 
emission). The CoM momenta of the pions 
were calculated assuming a two-body decay of the primary meson X, as is the case 
for the $\rho$. In cases where this assumption 
holds, the deviation from a 180$^{\circ}$ opening angle remains small and peaks at 0$^{\circ}$.
The data clearly show a large two-pion production
branch, which is rejected in the subsequent analysis as indicated by the dashed
vertical line. Only events with acoplanarities larger than 25$^{\circ}$ were
associated with three-body decays, $\pi$$^+$$\pi$$^-$Y. 
This criterion rejects more than 90\% of the two-body events while reducing the 
amount of $\omega$ events by only 30\%. Hence, it permits a clean identification of 
the $\omega$ production above a non-resonant background, mostly with multiplicities 
n$_{\pi}$~$\ge$~3 as shown in the missing mass plot of figure 2.
The apparent width of the resonance is dominated by the detector
resolution of $\sigma$~$\approx$~30~MeV at this energy. 
The $\eta$-meson also decays into three pions with 23\% probability. The cross
section for $\eta$-production at these energies is about five times larger than
that for $\omega$ production. Therefore, an appreciable amount of events in
the figure may be due to $\eta$ production. Because of the much lower mass of
the $\eta$ and the correspondingly higher available energy, the angular restrictions
on the proton directions and velocities reduce the acceptance for $\eta$ events to
1.9\% compared with 6.4\% for the $\omega$ and less than 0.5\% for the $\rho$ contribution 
to the spectrum in figure 2. Since the available energy is large, the protons which 
are detected in case of $\eta$ production will have large velocities. Therefore, 
the missing mass resolution for $\eta$ mesons is much worse than for the heavier and 
kinematically constrained $\omega$, so that an $\eta$ signal is not observed in 
figure 2. The emission direction of X was restricted to forward emission, 
$\theta$$_X ^*$~$\le$~90$^{\circ}$. This event pattern is also favored by the 
detector acceptance. It yields protons sufficiently 
low in energy to allow calculation of the missing mass with the quoted precision, 
while the backward emission of X boosts the protons to higher velocities so that the
missing-mass resolution degrades with increasing angle of X. 
The experimental mass spectrum (data points with statistical errors) exhibits 
a peak at the $\omega$ mass which is well reproduced in shape and position by the
detector Monte Carlo (hatched peak for the $\omega$) when added to a parametrized 
background, as shown by the shaded spectrum. The small remaining shift in
position is due to calibration uncertainties in the time-of-flight determination.
The missing mass peak of figure 2
can be integrated to yield the total number of $\omega$ mesons produced. Different
background shapes were applied in fits to the distribution. They produce cross 
sections for the $\omega$ which are consistent to within $\pm$10\%. The 
background used in producing figure 2 is taken to be a third order polynomial 
which has been fitted to the distribution above and below the peak in the indicated
region. While, in principle, a background estimation using simulations which incorporate 
non-resonant as well as resonant multi-pion production appears desirable, the lack of 
differential cross section data prevents doing so.\\
The normalisation of the data requires a luminosity calibration which is easily
accomplished in our experiment due to the simultaneous measurement of elastically 
scattered protons over a wide angular range. These data can then be compared to
the results of other experiments which measured the elastic scattering to a
high level of precision (see e.g. \cite{EDD}) and/or phase shift analyses
\cite{SAI}. The overall normalisation uncertainty from this procedure
is estimated to be about 5\%. After correction for the acceptance of the setup 
using Monte Carlo simulations, total cross sections can be deduced. They 
are shown in figure 3 together with results from the literature \cite{DIS00,Fla} and 
theoretical calculations \cite{Cas93,Sib96,Nak99}. The acceptance calculations
include detector resolution as well as all the conditions applied to the data. The 
overall acceptance for $\omega$ events at 173~MeV is 6.4\% with only a small 
modulation with angle (7.1\% at 92~MeV). A systematical uncertainty which is larger 
than the statistical errors comes about when an angular distribution of the 
$\omega$ mesons is taken into account in the acceptance calculations. To this end, 
the isotropic three-body phase space was modified in an iterative procedure 
to fit the observed anisotropy (cf. figure 4) at 173~MeV. This changes the 
deduced cross section by almost 20\%. Since the current analysis does not
permit the extraction of angular distributions at the lower energy, anisotropies 
cannot be excluded in this case. It is however reasonable to assume that these 
will not exceed those determined at twice the excess energy, so that the same
anisotropy was used for both energies to estimate the resulting change in cross section.
This prescription yields a decrease in cross section by 13\% (about half of the 
statistical error) at the lower energy and an increase by 19\% at the higher energy 
(almost twice the statistical error). This is caused by the different ranges of  
momenta accessible to the $\omega$ and protons at the two energies. In particular, 
the kinematic focussing of the pions from $\omega$ decay is very different in both 
cases. This changes the acceptance of the detector differently at the two energies. 
The displayed cross sections are $\sigma$~=~(7.5\,$\pm$\,1.9)\,$\mu$b at 
$\epsilon$~=~92~MeV and $\sigma$~=~(30.8\,$\pm$\,3.4)\,$\mu$b at 173~MeV, respectively,
where in the latter the angular distribution of the ejectiles was taken into 
account in the acceptance correction. The corrections to the data due to
background subtraction, acceptance calculations employing various conditions such as, 
e.g., using phase space distributions for the $\omega$ decay pattern instead of the
vector meson decay, do not alter the deduced cross sections by more than 20\% as
was also obtained from the anisotropy considerations. Thus, we conclude that the 
absolute numbers for the total cross sections given here are subject to systematic
uncertainties not larger than 20\% which are included in the error bars of figure 3.\\
Neither of the theoretical excitation functions based on parametrizations of 
one-pion exchange model calculations (dotted \cite{Cas93} and dash-dotted \cite{Sib96} 
lines in figure 3) is able to describe the measured data at low energy. The 
parametrization given by Sibirtsev \cite{Sib96} comes closer to the data than the 
one by Cassing \cite{Cas93} but fails below 90~MeV where final state interactions 
may start to contribute.\\
There is a general agreement among various theorists \cite{Nak99,Nak00,Sib00,Tit,Kai} that 
$\omega$ and $\phi$ production, when viewed in a meson exchange picture, can be ascribed 
to two dominant contributions which are called "nucleonic" when the meson is 
emitted by one of the nucleons (NN$\upsilon$ vertex, $\upsilon$ denotes either vector meson, 
$\phi$ or $\omega$) in contrast to the "mesonic" 
current where the emitted vector meson originates from the exchanged meson, e.g. from a 
$\rho$$\pi$$\upsilon$ vertex. The solid and dashed lines in figure 3 show two 
calculations by Nakayama et al. \cite{Nak99} within such a model. 
Different sets of parameters for both the nucleonic and mesonic currents were used.
The parameters for $\phi$ and $\omega$, in particular the coupling
constants, were related by SU(3) considerations. The form factor for the mesonic
current was fixed to the $\phi$ angular distribution at 85~MeV which is flat to within 
the -\,rather large\,- experimental errors \cite{DIS98}. The vector coupling constants for 
the $\rho$$\pi$$\upsilon$ vertices were taken from $\phi\,\rightarrow\,\rho \pi$ and
$\omega\,\rightarrow\,\gamma \pi$ decay, respectively, while the tensor-to-vector
ratio was varied within reasonable limits, $\kappa$$_{\upsilon}$\,=\,$\pm$0.5.
Left with the cutoff parameter for the nuclear contribution's form factor, the total 
pp$\omega$ cross sections \cite{Hib} were fitted. Finally, the NN$\phi$ coupling 
constant was adjusted to fit the nuclear contribution to the $\phi$ angular distribution.
While both calculations fail to reproduce our data, the solution with negative tensor
coupling, $\kappa$$_{\upsilon}$\,=\,-0.5 (solid line), gives a better agreement in overall 
shape. It uses a rather small cutoff parameter, $\Lambda$$_{N}$~=~1170~MeV, and results 
in a coupling constant of g$_{NN\phi}$\,=\,-0.45. Because of the decreasing nuclear 
contribution with decreasing $\Lambda$$_{N}$ and the destructive interference of vector 
and tensor currents, this solution results in an angular distribution which is completely 
dominated by the mesonic current. Hence, a flat distribution is expected, as is indeed 
observed in the case of $\phi$ production. However, the anisotropic pp$\omega$ angular 
distribution for an excess energy of 173~MeV which is shown in figure 4 suggests that 
this parameter set cannot describe the $\omega$ production. The figure does not include 
the most backward-emitted mesons because of the forementioned bias toward low-momentum
protons needed in the analysis. Since the entrance channel is symmetric 
the distribution has to be symmetric about $\cos \theta$$_{\omega} ^*$~=~0 as 
indicated by the two data points below $\cos \theta$$_{\omega} ^*$~=~0. The
angular distribution of protons was indeed found to be symmetric. The acceptance corrections
used to determine the differential cross sections were deduced separately for each interval 
in $\cos \theta$$_{\omega} ^*$. A decrease in acceptance from 12\% to 7\% is observed when going 
from $\cos \theta$$_{\omega} ^*$~=~0 to $\cos \theta$$_{\omega} ^*$~=~1, while the $\omega$
yield increases by a factor of two over the same range. Hence, the observed anisotropy 
is somewhat amplified by the acceptance correction, however the effect is already 
contained in the uncorrected data. The 
strong anisotropy which was also observed in \cite{DIS98} for 320~MeV 
excess energy suggests that higher partial waves may be present even at 
lower energy where, according to \cite{Nak98}, this may be viewed as direct 
proof of the dominance of nucleonic currents. A fit with the lowest two even
Legendre polynomials yields coefficients of 
a$_2$/a$_0$~=~2.5($\pm$0.3)\,/\,2.6($\pm$0.1) compared to 
3.1($\pm$0.2)\,/\,4.0($\pm$0.1) at 320~MeV \cite{DIS00} where, in addition, a P$_4$
contribution was needed which is not required here.\\
Using our $\omega$ cross section value at 92~MeV together with the
$\phi$ cross section \cite{DIS99} at 85~MeV yields a production ratio 
$\sigma$$_{\phi}$/$\sigma$$_{\omega}$ of about (3\,$\pm$\,1)\%. This is, although
smaller than the DISTO estimate without the current cross section,
about a factor of 7 larger than the OZI prediction.
The uncertainty is dominated by the 
experimental error on the $\phi$ cross section.
The current result also exceeds the prediction of about 1\% for the ratio at 
this energy in \cite{Nak00} for small NN$\phi$ coupling. More
detailed studies are however needed before final conclusions will emerge.\\
With envisaged improvements in experimental techniques, especially in accepted 
luminosity, also $\phi$ production may come into reach at COSY.
Further experimental investigations will then help to shed light on the question 
of the strange content of the nucleon adressed in vector meson production.\\

The authors would like to express their gratitude to the COSY staff for the 
operation of the accelerator during the experiment. Fruitful discussions with
C. Hanhart, K. Nakayama and J. Ritman are gratefully acknowledged.

\newpage
\begin{figure}[htb]
\epsfxsize=20cm
\centerline{
      \epsffile{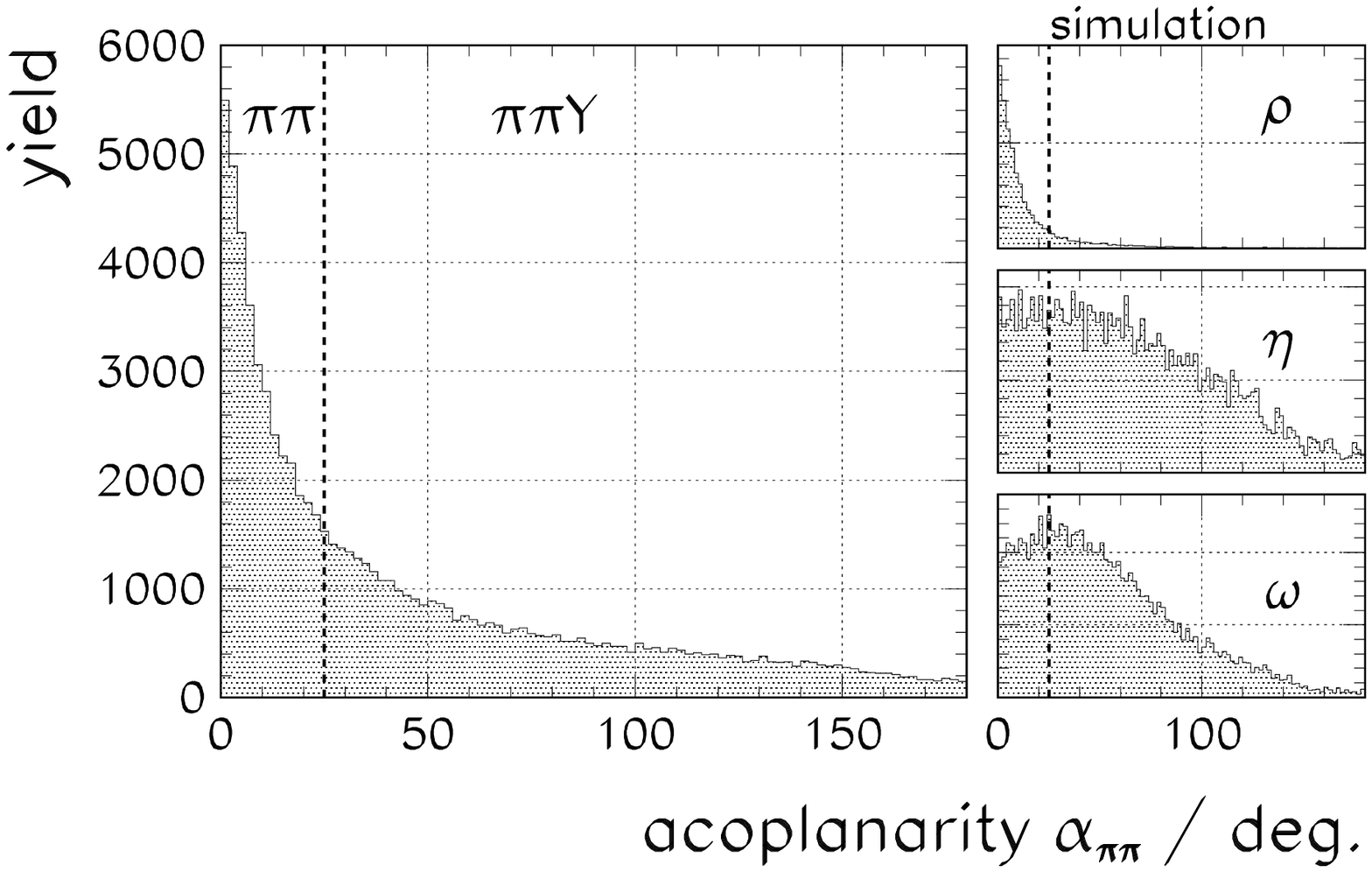}
           }
\vspace*{1mm}

  \caption{Distribution of acoplanarity angles derived using the reconstructed momentum
  of the hypothetical meson X and the directions of its decay pions. The large frame 
  shows the experimental spectrum which is used to separate two-pion events (labeled
  $\pi$$\pi$) to the left of the dashed line from those where an additional particle
  Y was involved. The small spectra to the right show the simulated angular distributions
  for the decays $\rho$$^0$\,$\rightarrow$\,$\pi$$^+$$\pi$$^-$ and the three-body decays
  $\eta$\,$\rightarrow$\,$\pi$$^+$$\pi$$^-$$\pi$$^0$ and  
  $\omega$\,$\rightarrow$\,$\pi$$^+$$\pi$$^-$$\pi$$^0$.}
\end{figure}

\newpage
\begin{figure}[htb]
\epsfxsize=20cm
\centerline{
      \epsffile{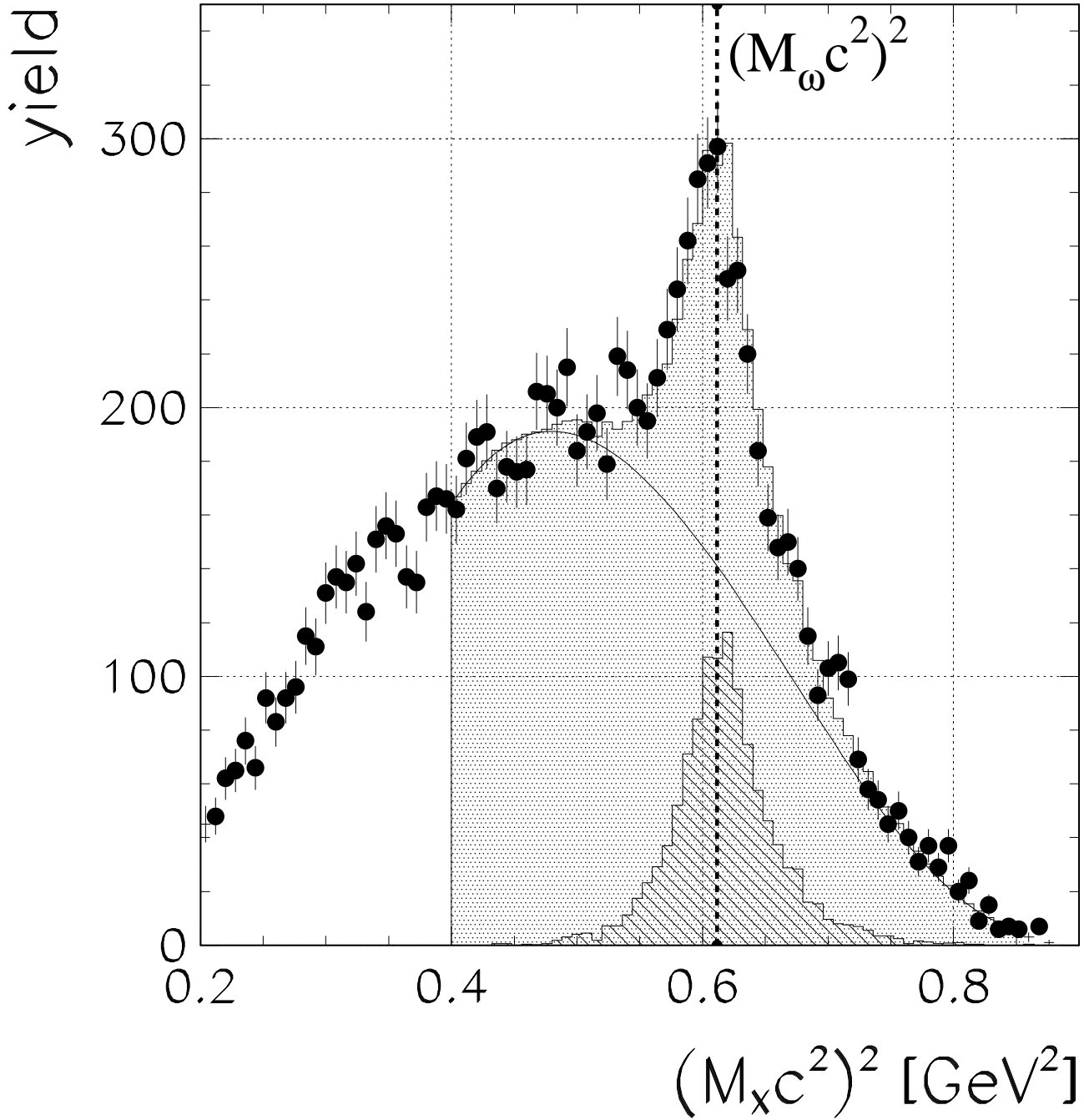}
           }
\vspace*{1mm}

  \caption{Spectrum of the squared missing mass as reconstructed from two detected 
  protons after enhancing cuts on the pions. The $\omega$ peak at $\approx$~0.6~GeV$^2$ is 
  clearly visible. The hatched peak at (Mc$^2$)$^2$~=~0.6~GeV$^2$ shows the simulated
  response function of the detector. It nicely agrees in shape with the $\omega$ signal 
  above background, as demonstrated by the shaded area which is a sum of the 
  parametrized background as indicated and the Monte Carlo $\omega$ signal (hatched peak).}
\end{figure}

\newpage
\begin{figure}[htb]
\epsfxsize=18cm
\centerline{
      \epsffile{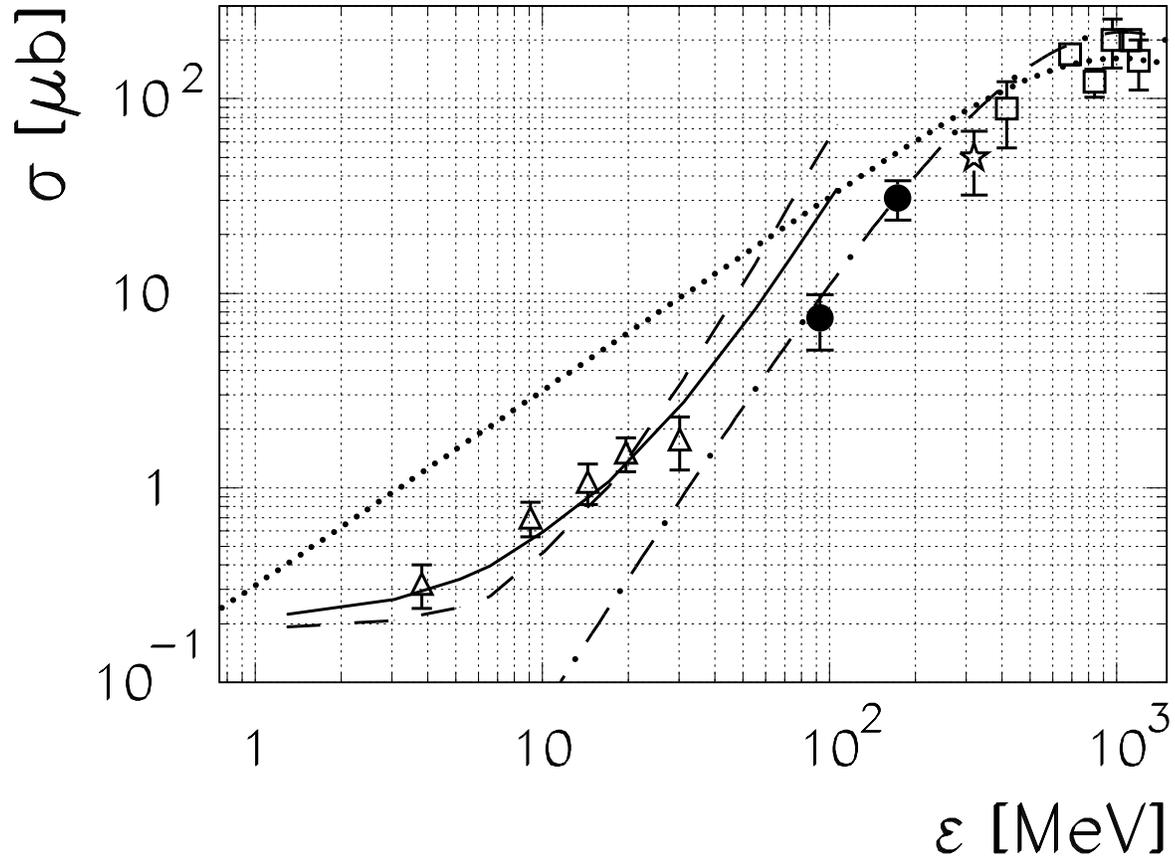}
           }
\vspace*{1mm}

  \caption{Excitation function of pp~$\rightarrow$~pp$\omega$. The lines are theoretical 
  calculations by Cassing (\cite{Cas93}, dotted line), Sibirtsev (\cite{Sib96}, 
  dash-dotted line) and Nakayama and coworkers (\cite{Nak99}, solid and dashed lines) 
  which are discussed in the text. The filled circles
  are the present data, while the open triangles are from the SPES3/Saturne 
  experiment \cite{Hib}, the star is from DISTO/Saturne \cite{DIS00}. The errors shown
  include systematic as well as statistical errors which were added quadratically.
  The open squares represent data adopted from a compilation of cross sections\cite{Fla}.}
\end{figure}

\newpage
\begin{figure}[htb]
\epsfxsize=20cm
\centerline{
      \epsffile{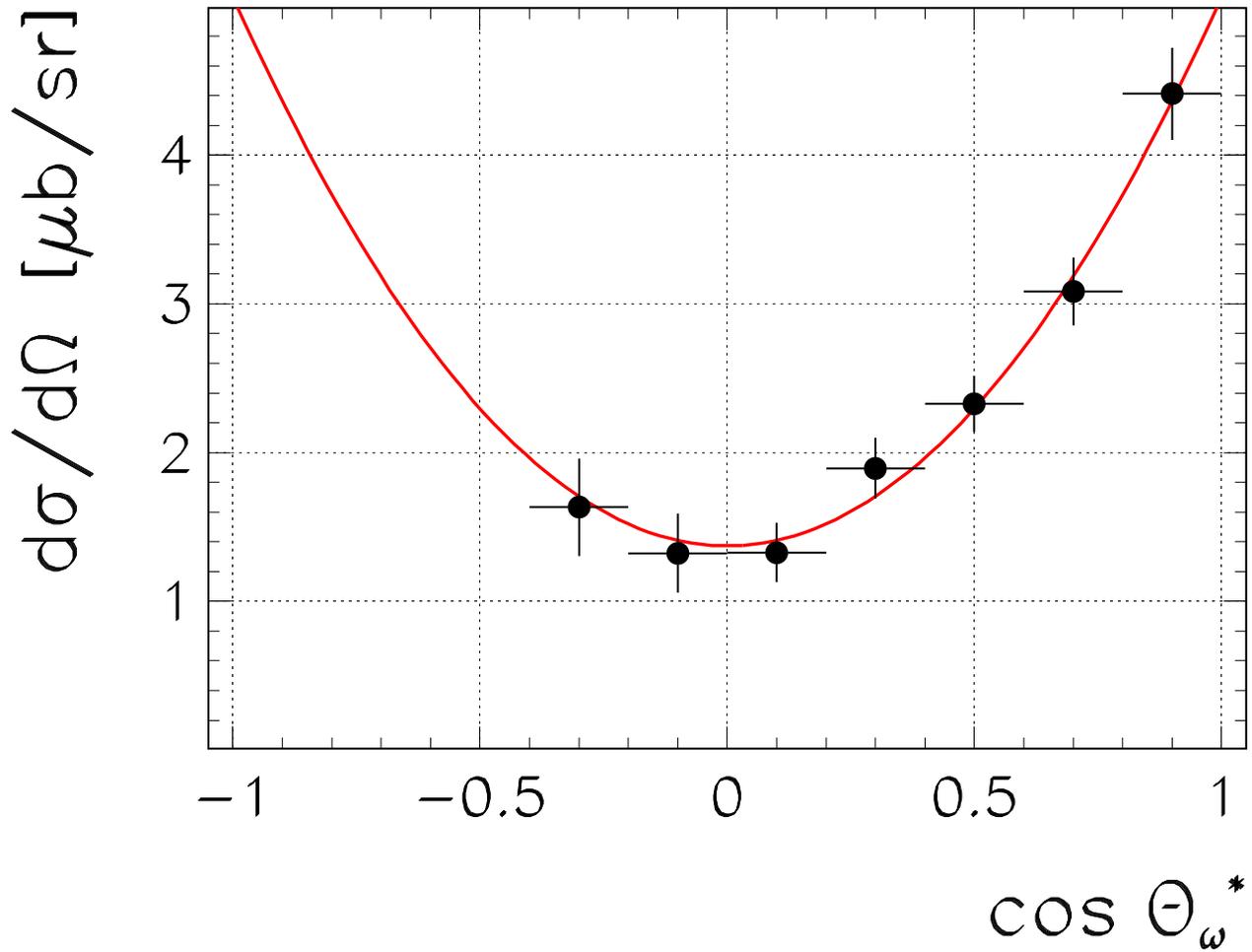}
           }
\vspace*{1mm}

  \caption{Angular distribution of the $\omega$ in the overall center-of-momentum 
  frame. Only the forward part, $\cos \theta$$_{\omega} ^*$~$\ge$~-0.3, is displayed since 
  the analysis is restricted to this region. The full line is a fit with the lowest two even 
  Legendre polynomials.}
\end{figure}

\end{document}